\documentclass[aps,nofootinbib,amsfonts,groupedaddress]{revtex4}
\usepackage{amsmath}
\usepackage{epsfig,epsf}
\usepackage{graphicx}
\usepackage{verbatim}

\def\beq{\begin{equation}}
\def\eeq{\end{equation}}
\def\bea{\begin{eqnarray}}
\def\eea{\end{eqnarray}}
\def\eq#1{{Eq.~(\ref{#1})}}
\def\fig#1{{Fig.~\ref{#1}}}

\newcommand{\bas}{\bar{\alpha}_s}

\newcommand{\Lb}{\left(}
\newcommand{\Rb}{\right)}
\newcommand{\nn}{\nonumber}

\def\pom{{I\!\!P}}

\newcommand{\h}{\frac{1}{2}}
\begin{document}

\title{Large $\mathbf{b}$ behaviour  in the CGC/saturation approach:  BFKL equation with pion loops}

\author{Eugene Levin}
\email{leving@post.tau.ac.il, eugeny.levin@usm.cl}
\affiliation{ Department of Particle Physics, School of Physics and Astronomy,
Tel Aviv University, Tel Aviv, 69978, Israel}
\affiliation{ Departamento de F\'\i sica,
Universidad T$\acute{e}$cnica Federico Santa Mar\'\i a   and
Centro Cient\'\i fico-Tecnol$\acute{o}$gico de Valpara\'\i so,
Casilla 110-V,  Valparaiso, Chile}

\date{\today}


\begin{abstract}
 In this paper we proposed a solution to the longstanding problem of the CGC/saturation approach: the power-like fall  of the scattering amplitudes at large $b$.  This decrease  leads to the violation of the Froissart theorem and  makes the approach theoretically inconsistent. We showed in the paper that sum of the pion loops  results in the exponential   fall of the scattering amplitude at large  impact parameters and in the restoration of the Froissart theorem.  
 \end{abstract}

\maketitle


\section{ Introduction}
It is well known that perturbative QCD  suffers  fundamental problem: the scattering amplitude falls down at large impact parameters ($b$) as a power of $b$. In particular  the CGC/saturation approach\cite{REV},  which is based on perturbative QCD, faces this problem. Such a power-like decrease leads to the violation\cite{KW,FIIM} of the Froissart theorem\cite{FROI}.   The violation of the Froissart theorem stems from the growth of the radius of interaction as a power of the energy; and 
 can be fixed by introducing a new dimensional scale in addition to the saturation momentum. Since even a power-like decrease provides the small values of the amplitudes at large $b$ , in the framework of the CGC/saturation approach we need to fix the large $b$ behaviour in the BFKL evolution equation\cite{BFKL,LIREV}. Therefore, we have to introduce two dimensional scales in the CGC/saturation approach: the saturation momentum, that is originated by the interactions of the BFKL Pomerons; and the new scale of the non-perturbative  source that provides the exponential decrease of the BFKL Pomeron at large $b$.
  
\begin{figure}[ht]
     \begin{center}
     \includegraphics[width=0.2\textwidth]{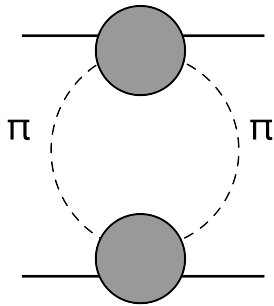} 
     \end{center}    
      \caption{ The exchange of two pions that gives the main contribution to the large $b$ behaviour of the scattering amplitude\protect\cite{FROI}. }
\label{2pi}
   \end{figure}

  The  problem of large $b$ behaviour of the BFKL Pomeron has not been solved in spite of the numerous  attempts based both on analytical and numerical calculations, to approach it \cite{FIIM, LERYB1,LERYB2,GBS1,GKLMN,HAMU,MUMU,BEST1,BEST2,KOLE,LETA,LLS}. We need to find the sincere  theoretical way to introduce the second dimensional scale. The proof of the Froissart theorem indicates that the exponential fall down at large $b$ is closely related to the contribution of the exchange of two lightest hadrons: pion, in $t$-channel (see \fig{2pi}).  
    
\begin{figure}[ht]
     \begin{center}
     \includegraphics[width=0.4\textwidth]{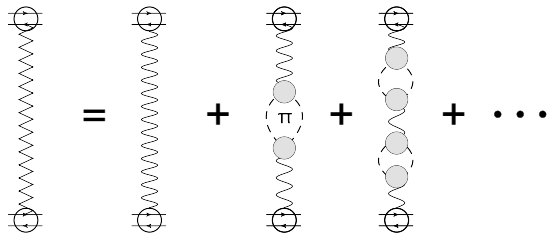} 
     \end{center}    
      \caption{The diagrams that describe the pion loop contribution to the BFKL Pomeron. The wavy line describe the BFKL Pomeron, the dashed line stands for pion. The zigzag line denotes the resulting Pomeron.  }
\label{piloop}
   \end{figure}

 In this paper we propose the theoretical approach which based  on the assumption that the BFKL Pomeron, being perturbative in nature, takes into account rather short distances (say of the order of $1/m_G$, where $m_G$ is the mass of the lightest glueball); and the long distance contribution can be described by the exchange of pions ($1/2\mu\,\gg\,1/m_G$, where $\mu$ is the pion mass).
 We modify the BFKL Pomeron by including the pions loops shown in \fig{piloop} and show that the resulting Pomeron leads to  an exponentially small scattering amplitude at large $b$. Such behaviour of the scattering amplitude heals all difficulties of the perturbative approach including the violation  of the Froissart theorem.

 \section{The BFKL Pomeron}
 In this section we are going to describe that main features of the BFKL Pomeron which we will need below.   All these features are known and can be found in Refs.\cite{BFKL,LIREV,LIPB}. We include them for the sake of completeness and of clarification of   notations. Let us consider the scattering amplitude of two dipoles with sizes $r_1$ and $r_2$ at high energy $s$( $Y = \ln s$). The Green's function of the BFKL Pomeron to the  scattering amplitude at fixed momentum transferred along the Pomeron:$Q_T$ takes the following form\cite{LIPB}
 \beq \label{BFKL}
\mbox{Im}\, G_{\pom}\Lb Y, t = - Q^2_T; r_1, r_2\Rb\,\,=\,\,r_1\,r_2\,\int^{\epsilon + i \infty}_{\epsilon - i \infty}\frac{d \omega}{ 2 \pi i}\,e^{\omega\,Y}\,\,G_{\pom}\Lb \omega, Q_T; r_1, r_2\Rb
 \eeq
 
 The integration contour is situated to the right of all singularities of $G_\pom$ in $\omega$. 
 Function $G_\pom\Lb \omega, Q_T; r_1, r_2\Rb$ has been found in Ref.\cite{LIPB} and takes the following form (see Eq.35 in Ref.\cite{LIPB})
 \beq \label{GBFKL}
 G_\pom\Lb \omega, Q_T; r_1, r_2\Rb\Big{|}_{\begin{minipage}{1.5cm}{$\omega \to \omega_0\\ 
 Q_T \to 0$ }\end{minipage}}\,\,=\,\,\,\frac{\pi}{\kappa_0}\Bigg\{ \Lb \frac{r_1}{r_2}\Rb^{\kappa_0}\,\,+\,\, \Lb \frac{r_2}{r_1}\Rb^{\kappa_0}\,\,-\,\,2\,\Lb Q^2_T\,r_1\,r_2\Rb^{\kappa_0}\Bigg\}
  \eeq 
  
  where $\kappa_0$ is the solution of the following equation
\beq\label{CHI}
\omega\,=\,2\,\bar{\alpha}_S \left( \psi(1) - Re{\, \psi\left(\frac{1}{2} -
\kappa_0\right)}\right)\,\,=\,\,\omega_0\,\,+\,\,D\,\kappa^2_0 \,\,+\,\,{\cal O}\Lb \kappa_0^3\Rb
 \eeq
 and it is equal to
 \beq \label{KAPPA}
\kappa_0\,\,=\,\,\sqrt{\frac{\Lb \omega\,\,-\,\,\omega_0\Rb}{ D}}~~~~~~\mbox{with}~~~
\omega_0\,=\,\bas 4 \ln 2~~\mbox{and}~~D\,=\,\bas 14 \zeta\Lb 3\Rb
\eeq

From \eq{GBFKL} one can see that in the kinematic region  of small $Q_T$ where $|(\omega - \omega_0)\,\ln^2\Lb Q^2_T r_1 r_2\Rb|\,\gg\,1$  $  G_\pom\Lb \omega, Q_T; r_1, r_2\Rb\,\,\to\,\, 1/\sqrt{\omega - \omega_0}$ while ion the region of $|(\omega - \omega_0)\,\ln^2\Lb Q^2_T r_1 r_2\Rb|\,\sim\,1$ $G_\pom$ is less singular ($G_\pom\Lb \omega, Q_T; r_1, r_2\Rb\,\,\to\,\, \sqrt{\omega - \omega_0}$.

\begin{figure}[ht]
     \begin{center}
     \includegraphics[width=0.6\textwidth]{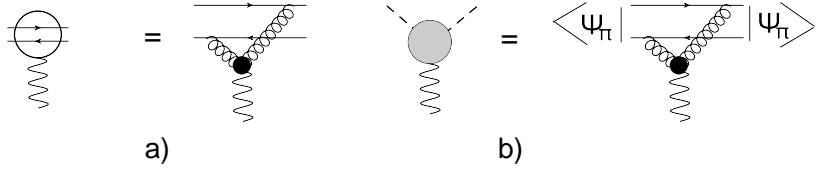} 
     \end{center}    
      \caption{ The graphic illustration of \protect\eq{SCA}: for dipole-Pomeron vertex (\fig{ver}-a ) and for the vertex of the interaction of the BFKL Pomeron with a pion (\fig{ver}-b). }
\label{ver}
   \end{figure}

  The scattering amplitude is equal to
  \beq \label{SCA}
  A\Lb Y, Q_T\Rb\,\,=\,\,\int d^2 r'_1 \,\int d^2 r'_2\,\, |\Psi\Lb r'_1\Rb|^2  \,|\Psi\Lb r'_2\Rb|^2  \,
  \mbox{Im}\, G_{\pom}\Lb Y, t = - Q^2_T; r'_1, r'_2\Rb
  \eeq
  where $\Psi$ denotes the wave function of the scattering particle (dipole). For the scattering of two dipoles with the sizes $r_1$ and $r_2$ $|\Psi_{dip}\Lb r'_1\Rb|^2 $ takes the form
  $    |\Psi_{dip}\Lb r'_1\Rb|^2 = 2 \bas \delta\Lb \vec{r} - \vec{r}^{\,'}\Rb$ (see \fig{ver}-a).
  For the scattering of pion $\Psi_\pi\Lb r_1\Rb$ is the non-perturbative wave function of the $\pi$-meson (see \fig{ver}-b).
\section{Summing pion loops}
\subsection{General solution}
\begin{figure}[ht]
     \begin{center}
     \includegraphics[width=0.7\textwidth]{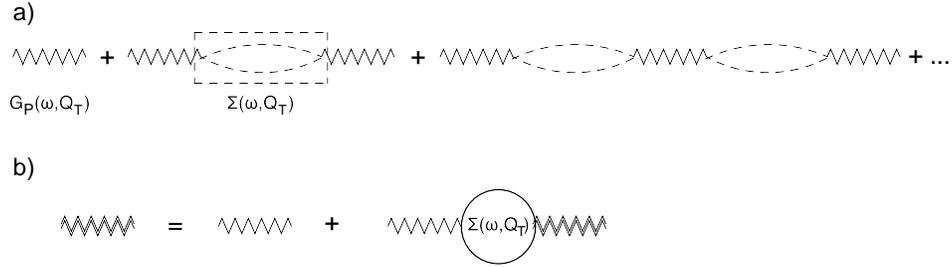} 
     \end{center}    
      \caption{ The diagrams of \fig{piloop} in $\omega$-representation (\protect\fig{eq}-a) and the graphic form of the equation that sums these diagrams (\protect\fig{eq}-a) }
      \label{eq}
   \end{figure}

The variables $\omega$ and $Q_T$ are convenient to sum diagrams of \fig{piloop} since both variables  are preserve  along the diagrams of \fig{piloop}. Indeed, the summation of these diagrams can be done easily (see \fig{eq}). The equation that sums the diagrams of \fig{eq}-a  can be written in the following analytical form:
\beq \label{EQ}
\mathbf{G}_\pom\Lb \omega, Q_T; R_\pi,R_\pi\Rb \,\,=\,\,G_\pom\Lb \omega, Q_T; R_\pi,R_\pi\Rb\,\,\,+\,\,G_\pom\Lb \omega, Q_T; R_\pi,R_\pi\Rb \,\Sigma\Lb \omega, Q_T\Rb
\mathbf{G}_\pom\Lb \omega, Q_T; R_\pi,R_\pi\Rb
\eeq
where $\mathbf{G}_\pom$ denotes the resulting Green's function of the Pomeron with the pion loops for $r_1=r_2 = R_\pi$ where $R_\pi$ is the size of the pion. In other words , it is typical size of the integral of \eq{SCA} for pion scattering (see \fig{ver}-b).

Solution to \eq{EQ} takes the form

\beq \label{SOL}
\mathbf{G}_\pom\Lb \omega, Q_T; R_\pi,R_\pi\Rb \,\,=\,\,\frac{1}{G^{-1}_\pom\Lb \omega, Q_T; R_\pi,R_\pi\Rb\,\,-\,\,
\Sigma\Lb \omega, Q_T\Rb}
\eeq

We use the $t$-channel unitarity for $\mathbf{G}_\pom\Lb \omega, Q_T; R_\pi,R_\pi\Rb$ to calculate $\Sigma$,  as it was suggested in Ref.\cite{ANGR}.  The unitarity constraints for $\mathbf{G}_\pom$ takes the form \cite{COL,GRLEC} ($t = - Q^2_T$):
\beq \label{UNIT}
-i \{ \mathbf{G}_\pom\Lb \omega, t + i\epsilon ; R_\pi, R_\pi\Rb \,\,-\,\,\mathbf{G}_\pom\Lb \omega, t - i\epsilon ; R_\pi, R_\pi\Rb  \}\,\,=\,\,\rho\Lb \omega, t \Rb \mathbf{G}_\pom\Lb \omega, t + i\epsilon ; R_\pi, R_\pi\Rb\,\mathbf{G}_\pom\Lb \omega, t - i\epsilon ; R_\pi, R_\pi\Rb
\eeq
\eq{UNIT} is written for $ t \,>\, 4 \mu^2$ where $\mu$ is the mass of pion. We wish to sum only two pions contribution
assuming that all other can be included and have been taken into account in the BFKL Pomeron. Therefore, we believe that \eq{UNIT} can be trusted for $t \,<\,m^2_G$ where $m_G$ is the lightest glueball. $\rho \Lb \omega, t\Rb$ is equal to $( t - 4 \mu^2)^{\frac{3}{2} + \omega}/\sqrt{t}$.
\subsection{$\mathbf{\Sigma}$}
Plugging in \eq{UNIT} the solution of \eq{SOL} one can see that we obtain the following equation for $\Sigma$:
\beq \label{EQSIG}
\mbox{Im}_t \Sigma\Lb \omega,t\Rb\,\,=\,\,\h \rho \Lb \omega, t\Rb\,\Big(\Gamma^\pom_{\pi \pi}\Big)^2\,\frac{1}{\omega + 2}
\eeq
where vertex $\Gamma^\pom_{\pi \pi}$ describes the interaction of the BFKL Pomeron with $\pi$-meson (see \fig{ver}-b) and takes the simple form (see \eq{SCA} and \fig{ver}-b). Factor $1/(\omega + 2)$ arises in the angular momentum representation from the behaviour of the scattering amplitude due to exchange of two pions as $1/s$,

\beq \label{POMPIPI}
\Gamma^\pom_{\pi \pi}\,\,=\,\,\frac{8}{9}\bas \int d^2 r \,  r\, |\Psi_\pi \Lb r \Rb|^2\,\,=\,\,\frac{8}{9}\,\bas\,R_\pi
\eeq
Factor $8/9$ includes the colour factor and sum of two diagrams in \fig{ver}-b.

Experimentally $< R^2_\pi> = 0.438 \pm 0.008 \,fm^2$ \cite{PIR} and this value we will bear in mind  for the numerical estimates.

We use the dispersion relations to calculate $\Sigma$. Since we believe that non-perturbative corrections will not change the intercept of the BFKL Pomeron (see Refs.\cite{LETA,LLS}) we write the dispersion relation with one subtraction. It takes the form
\beq \label{DISRE1}
\Sigma\Lb \omega,t\Rb\,\,=\,\, t\, \frac{1}{\pi} \int \frac{ d t' \,\mbox{Im}_t \Sigma\Lb \omega,t'\Rb}{ t' \,(t' - t)}\,\,=\,\,t\frac{3}{2}\,\frac{1}{\omega + 2}\frac{\Lb \Gamma^\pom_{\pi \pi} \Rb^2}{  \pi} \int_{4 \mu^2}^{m_G^2} \frac{d t'  \,\rho \Lb \omega, t'\Rb}{ t' \,(t' - t)}\,\,=\,\,\bas^2\,\frac{16}{9}\,R^2_\pi\,\,\frac{1}{\omega + 2}\,\,t\int_{4 \mu^2}^{m_G^2} \,\frac{d t'  \,\rho \Lb \omega, t'\Rb}{ t' \,(t' - t)}\eeq
Factor $3/2$ stems from two states in $t$-channel: $\pi^+\,\pi^-$ and $\pi^0\,\pi^0$.

The integral over $t'$ is divergent even after one subtraction. We evaluate this integral making the physical assumption that $\mbox{Im}_t \Sigma\Lb \omega,t'\Rb$ is small for $t' > m^2_G$ since  all singularities for larger $t'$ has been taken into account in the BFKL Pomeron contribution.
The integral over $t'$ for $\omega\, \ll\,1$ has been taken in Ref.\cite{ANGR}. For fixed $\omega$
\beq \label{DISRE2}
\Sigma\Lb \omega,Q_T\Rb\,\,\,=\,\,\,\bas^2\,\frac{16}{9}\,R^2_\pi\,\,\frac{1}{\omega + 2}\,\,(- Q^2_T)\,\frac{\Lb m^2_G\Rb^{5/2 + \omega}}{ 8 \mu^2 *\Lb 5 + 2 \omega\Rb \Lb 4 \mu^2 + Q^2_T\Rb}\,F_1\Lb \frac{5}{2} + \omega, \frac{3}{2}, 1, \frac{7}{2} + \omega, -\frac{m^2_G}{4 \mu^2}, -\frac{m^2_G}{4 \mu^2 + Q^2_T}\Rb
\eeq
where $F_1$ is the Appell $F_1$ function ( see Ref.\cite{RY} formulae {\bf 9.180 - 9.184}).

The linear term in $Q^2_T$ can be easily found since two arguments in $F_1$ coincide and we can use the relation (see Ref.\cite{RY} formula {\bf 9.182(11)})  and reduce \eq{DISRE2} to the form
\beq \label{DISRE3}
\Sigma\Lb \omega,Q_T\Rb\,\,\,=\,\,\,\bas^2\,\frac{16}{9}\,R^2_\pi\,\,\frac{1}{\omega + 2}\,\,(- Q^2_T)\,\frac{\Lb m^2_G\Rb^{5/2 + \omega}}{ 8 \mu^2 *\Lb 5 + 2 \omega\Rb \Lb 4 \mu^2 + Q^2_T\Rb}\,{}_2F_1\Lb \frac{5}{2} + \omega, \frac{5}{2},  \frac{7}{2} + \omega, -\frac{m^2_G}{4 \mu^2}\Rb
\eeq
For $m_g \,\gg\,\mu$ \eq{DISRE3} takes the form
\bea \label{DISRE4}
\Sigma\Lb \omega,Q_T\Rb&=&\,\,\,\bas^2\,\frac{16}{9}\,R^2_\pi\,\,\frac{1}{\omega + 2}\,\,(- Q^2_T)\\
&\times&\Bigg\{\frac{1}{\sqrt{\pi}\, 3 \Lb 5 + 2 \omega\Rb}
4^{1+\omega} \Lb \frac{1}{\mu^2}\Rb^{\h + \omega}\!\!\!\!\!\!\mu\, \Lb 2 + 3 \omega + \omega^3\Rb \Gamma\Lb - 2 - \omega\Rb \Gamma\Lb \frac{7}{2} + \omega\Rb\,
+\,\frac{\Lb m^2_G\Rb^\omega}{\omega}\,\frac{4 + 2 \omega + 3 \omega^2 + \omega^3}{4 ( 2 + \omega)}\Bigg\}\nn\\
&=& -Q^2_T \, \alpha'_{eff}\Lb \bas\Rb\nn
\eea

\begin{figure}[ht]
     \begin{center}
     \includegraphics[width=0.4 \textwidth]{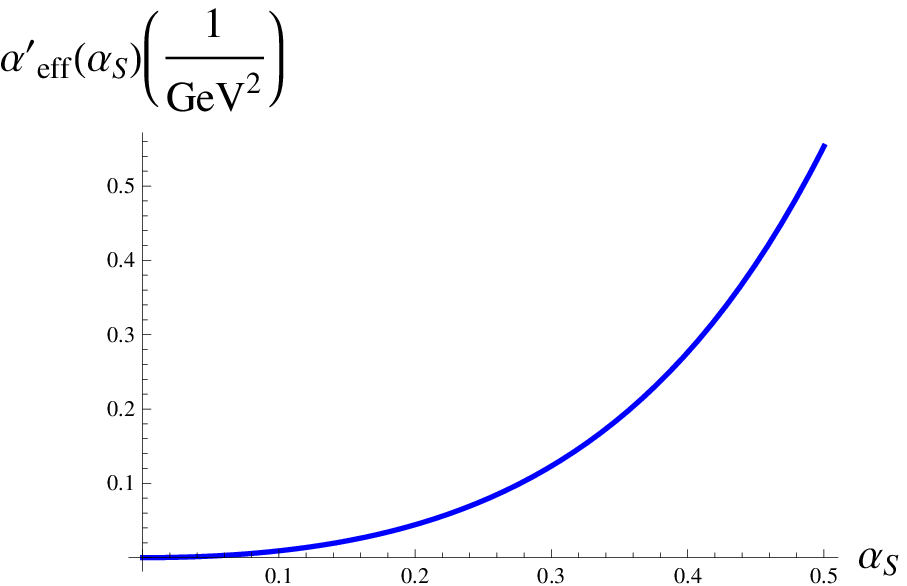} 
     \end{center}    
      \caption{ Function $\alpha'_{eff}\Lb \bas\Rb $ of \protect\eq{DISRE4} versus $\bas$, $ \omega = \omega_0 = 4 \ln 2 \bas$. }
      \label{sl}
   \end{figure}

We evaluate  $\alpha'_{eff}\Lb \bas\Rb$ using $R^2_\pi = 0.436  fm^2$ and 
the value for $m^2_G = 5 \,GeV^2$ that stems from lattice calculation of the glueball masses \cite{GLUETR} for the Pomeron trajectory.
 $\alpha'_{eff}\Lb \bas\Rb$ is shown in \fig{sl}.   In principle we can consider this coefficient as the only non-perturbative input that we need, but it is pleasant to realize that simple estimates give a sizable value.

      \subsection{Impact parameter dependence of the resulting Pomeron Green's function}
      Plugging $\Sigma$ from \eq{DISRE4} into \eq{SOL} we obtain that
      \beq \label{SOLMR}
    \mathbf{G}_\pom\Lb \omega, Q_T; R_\pi,R_\pi\Rb \,\,=\,\,\frac{\pi \sqrt{D}}{\sqrt{\omega\,-\,\omega_0}\,\,+\,\,\alpha'_{eff}\pi \sqrt{D}\,Q^2_T}
    \eeq  
  Calculating  $   \mbox{Im}\, G_{\pom}\Lb Y, t = - Q^2_T; r_1, r_2\Rb$ we substitute \eq{SOLMR} into \eq{BFKL} and obtain the following equation
  \beq \label{SOLY}
   \mbox{Im}\, G_{\pom}\Lb Y, Q_T; R_\pi, R_\pi\Rb =  R^2_\pi
   \int_{C} \frac{d \omega}{ 2 \pi i}\,e^{\omega\,Y}\,\,G_{\pom}\Lb \omega, Q_T; R_\pi,R_\pi\Rb =
  \int_C\!\frac{d\, \delta\omega}{2 \pi i}\,e^{\omega_0\,Y} \, e^{ \delta\omega\,Y} \frac{\pi \sqrt{D}R^2_\pi}{\sqrt{\delta\omega}\,\,+\,\,\alpha'_{eff}\pi \sqrt{D}\,Q^2_T}
     \eeq
 where $\delta\omega = \omega \,-\,\omega_0$.  
     \begin{figure}[ht]
     \begin{center}
     \includegraphics[width=0.2 \textwidth]{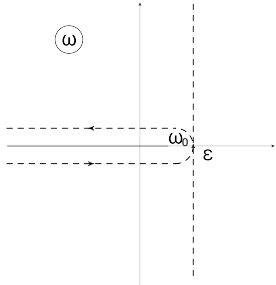} 
     \end{center}    
      \caption{ The contour of integration over $\omega$ in \protect\eq{SOLY}. }
      \label{cont}
   \end{figure}
      
      In the impact parameter representation \eq{SOLY} takes the form
      \bea \label{SOLB}
    \mbox{Im}\, G_{\pom}\Lb Y, b; R_\pi, R_\pi\Rb& = & 2 \pi\,R^2_\pi\, \int^\infty_0  Q_T d Q_T J_0\Lb Q_T b\Rb \int_C\!\frac{d\,\delta\omega}{2 \pi i}\,e^{\omega_0\,Y} \, e^{ \delta\omega\,Y} \frac{\pi \sqrt{D}}{\sqrt{\omega\,-\,\omega_0}\,\,+\,\,\alpha'_{eff}\pi \sqrt{D}\,Q^2_T}
 \nn\\
    &=&   \frac{2 \pi R^2_\pi}{\alpha'_{eff}}\, \int_C\!\!\! \delta\omega \,e^{\omega_0\,Y} \, e^{ \delta\omega\,Y}\, \,K_0\Lb \sqrt{\frac{\sqrt{\delta \omega}}{\alpha'_{eff} \,D\,\pi}}\,\, b\Rb      \eea
    $K_0$ in \eq{SOLB} is modified Bessel function of the second kind (McDonald function) (see Ref.\cite{RY} formula   {\bf 8.43}). It is needed in \eq{SOLB} to take integral over $\delta\omega$ which is rather difficult to perform in a general form. However, as we will show below, the extended information on physical observables can be extract from \eq{SOLB} and \eq{SOLY} using different asymptotic expantions for which we will be able to take the integral over $\delta \omega$.
    
    \section{Physical observables}
    \subsection{ The shrinkage of the diffraction peak.}
    The diffraction slope  $B$ is defined as
    \beq \label{SDP1}
    \frac{d \sigma_{el}}{d Q^2_T}\,\,=\,\,\frac{d \sigma_{el}}{d Q^2_T}\Big{|}_{Q_T=0}\,e^{-B\,Q^2_T}
    \eeq
    where $\sigma_{el} $ is the elastic cross section.
    
   Since $d \sigma_{el}/d Q^2_T \propto |A|^2$ where $A$ is the scattering amplitude, one can see that $B \,\,= 2 d \ln A/dQ^2_T $ at $Q_T=0$.    Using \eq{SOLY} we can calculate the slope: viz.
   \bea \label{SDP2}
   B\,&=&\, 2 \frac{ d \mbox{Im}\, G_{\pom}\Lb Y, Q_T; R_\pi, R_\pi\Rb \Big{/} d Q^2_T|_{Q_T=0}}{ \mbox{Im}\, G_{\pom}\Lb Y, Q_T; R_\pi, R_\pi\Rb|_{Q_T=0}}\,\,=\,\,2\,\frac{ \alpha'_{eff} \sqrt{D} \pi \int_C\!\frac{d\, \delta\omega}{2 \pi i}\,e^{\omega_0\,Y} \, e^{ \delta\omega\,Y} \frac{\pi \sqrt{D}}{\sqrt{\omega\,-\,\omega_0}}}{   \int_C\!\frac{d\, \delta\omega}{2 \pi i}\,e^{\omega_0\,Y} \, e^{ \delta\omega\,Y} \frac{\pi \sqrt{D}}{\omega\,-\,\omega_0} } \nn\\
   \,&=&\,2 \,\alpha'_{eff}\,\sqrt{D}\, \sqrt{Y}\,\equiv\,2 \,\alpha'_{eff}\,\sqrt{D}\, \sqrt{\ln s}\eea
   Integrating over $\delta \omega$ in \eq{SDP2} we close contour $C$ as it is shown in \fig{cont}.

$B \propto \sqrt{\ln s}$ is quite different from the Regge behaviour of $B \propto \ln s$ but it has been expected \cite{LERYB2}. The slope $B$ is related to average impact parameter in the interaction $B = \h\langle b^2 \rangle$ and its increase with energy occurs due to Gribov's diffusion in the transverse plane. At each emission the parton changes its position in $b$ by $\Delta b \propto 1/p_T$ where $p_T$ is the transverse momentum of the parton. In the BFKL Pomeron the average transverse momentum  increases with energy faster than $\ln s$ . Therefore, $\Delta b \to 0$ and we do not see any dependence of $\langle b^2 \rangle$ with energy for the BFKL Pomeron. Such a dependence is closely related to the confinement of parton in QCD. As it was argued in Ref.\cite{LERYB2} if we assume that all transverse momenta of partons due to confinement are larger than some dimensional parameter $\mu$($p_T \geq \mu$), in this case all parton with $p_T \sim \mu$ can have Gribov's diffusion with $\langle b^2 \rangle \propto \frac{1}{\mu^2 }\,\ln s$ . However, to calculate the average slope we need to multiply this  $  \langle b^2 \rangle$ by the probability to find a parton ($P(p_T = \mu)$) with $p_T \sim \mu$ which is $\propto 1/\sqrt{\ln s}$ . Finally 
$B \propto \,P(p_T = \mu)  \frac{1}{\mu^2} \,\ln s \propto  \frac{1}{\mu^2}\,\sqrt{\ln s}$.

Therefore, our approach gives the possibility to incorporate these physical ideas and suggests the theoretical procedure to introduce a new dimensional scale of the non-perturbative origin. It also allows to  estimate the value of this scale. 

The typical values of $Q^2_T$ for which we can trust \eq{SDP1}  can be estimated from $ B\,Q^2_T \,\leq\,1$ or $Q^2_T\,\leq \,1/\Lb 2 \,\alpha'_{eff}\,\sqrt{D}\, \sqrt{\ln s}\Rb \,\,\ll\, 1/R^2_\pi$. Such small values of $Q_T$ justifies the use of \eq{GBFKL} at small $Q_T \to 0$.

\subsection{ Large impact parameter behaviour of the scattering amplitude}
We can use the asymptotic behaviour of  $K_0$ at large $b$ and reduce integral of \eq{SOLB} to the   following expression
\beq \label{LB1}
    \mbox{Im}\, G_{\pom}\Lb Y, b; R_\pi, R_\pi\Rb\,\,\xrightarrow{b^2 \gg \alpha'_{eff}}\,\,  \frac{2 \pi\,R^2_\pi}{\alpha'_{eff}}\, \int_C\!\!\! \delta\omega \,e^{\omega_0\,Y} \, e^{ \delta\omega\,Y}\, \,\exp\Lb -  \sqrt{\frac{\sqrt{\delta \omega}}{\alpha'_{eff} \,D\,\pi}}\,\, b\Rb \eeq
    
    The integral over $\delta \omega$ can be taken using the steepest decent method. The equation for the saddle point in $\delta \omega$ takes the form
    \beq \label{LB2}
    Y\,-\,\frac{1}{4} \Lb \Lb\delta \omega_{SP}\Rb^{-\frac{3}{4} }\sqrt{\frac{ 1}{\alpha'_{eff} \,D\,\pi}}\,\,\,b\,\, \Rb\,\,=\,\,0;~~~~~~~~~~~~~~~\delta \omega_{SP}\,\,=\,\,\Lb \frac{ b}{4\, \sqrt{\alpha'_{eff} \,D\,\pi} \,Y}\Rb^{\frac{4}{3}};
      \eeq
    
    Plugging  \eq{LB2} in \eq{LB1} we obtain that
    \beq \label{LB3}
      \mbox{Im}\, G_{\pom}\Lb Y, b; R_\pi, R_\pi\Rb    \,\,\,\propto\,\,\,e^{\omega_0 \,Y}\,\exp\Lb - 3  \Lb \frac{b^4}{ 256 \,\alpha'^2_{eff}\, D^2\, \pi^2\, Y}\Rb^{\frac{1}{3}}\Rb
      \eeq
      
      From \eq{LB3} we see that scattering amplitude falls down exponentially. As we will show below such behaviour will restore the Froissart theorem and  transforms the CGC/saturation approach into self-consistent effective theory describing QCD at high energies.
      \subsection{Restoration of the Froissart theorem}
       The total cross section $\sigma_{tot}$ in our approach is equal to
       \beq \label{SIG1}
         \sigma_{tot}\\,=\,\,2 \int d^2 b\, \mbox{Im} A\Lb s, b\Rb \,\,=\,\,2 \int^{b_0}_0 d ^2 b\,\mbox{Im} A\Lb s, b\Rb  \,\,+\,\,2 \int^{\infty}_{b_0 }d ^2 b\,\mbox{Im} A\Lb s, b\Rb 
         \eeq
Following Ref.\cite{FROI} we estimate the behaviour of the total cross section dividing the integration over the impact parameter in two parts : $b \leq b_0$, where amplitude   $    \mbox{Im} A\Lb s, b\Rb\,\leq\,1$    due to $s$-channel unitarity ; and $b \geq b_0$ where  $ \mbox{Im} A\Lb s, b\Rb      \ll 1$. The value of $b_0$ can be estimated from the condition that $\mbox{Im} A\Lb s, b\Rb\,=\,f \,\ll\,1$.   In our approach this condition can be re-written, using \eq{LB3},  as follow
\beq \label{SIG2}
\mbox{Im} A\Lb s, b\Rb\,\propto\,    \mbox{Im}\, G_{\pom}\Lb Y, b; R_\pi, R_\pi\Rb\,\xrightarrow{ b^2 \gg\,\alpha'_{eff}}\,e^{\omega_0 \,Y}\,\exp\Lb - 3  \Lb \frac{b^4}{ 256 \,\alpha'^2_{eff}\, D^2\, \pi^2\, Y}\Rb^{\frac{1}{3}}\Rb\,\,=\,\,f
\eeq
\eq{SIG2} leads to 
\beq \label{SIG3}
   b_0\,\,=\,\,\frac{4}{3} \omega^{4/3}_0  \sqrt{\alpha'_{eff} \,D\,\pi} \,\,\,Y
   \eeq
   Plugging this $b_0$ into \eq{SIG1} we obtain
   
   \beq \label{SIG4}
   \sigma_{tot}\,\,\leq\,\,4 \pi b^2_0\,\,+\,\,4\, \pi\, f\, \frac{16\,D\,\pi}{3 \sqrt{3}}\,\Lb \alpha'_{eff} \,b_0\Rb^{2/3}\,\,\sim \,\,Y^2 = \ln^2 s
   \eeq
   Therefore, we see that the Froissart theorem is correct  in our approach.
   \subsection{$\mathbf{Q_T}$ dependece of the scattering amplitude.}
   Integral over $\omega$ in \eq{SOLY} can be taken if we close contour $C$ as it is shown in \fig{cont}. \eq{SOLY} takes the form
   \bea \label{QT1}
 &&  \mbox{Im}\, G_{\pom}\Lb Y, Q_T; R_\pi, R_\pi\Rb~ =~ R^2_\pi\,
  \int_C\!\frac{d\, \delta\omega}{2 \pi i}\,e^{\omega_0\,Y} \, e^{ \delta\omega\,Y} \frac{\pi \sqrt{D}}{\sqrt{\delta\omega}\,\,+\,\,\alpha'_{eff}\pi \sqrt{D}\,Q^2_T}\,\,=\,\, \sqrt{D}\,R^2_\pi\,e^{\omega_0 \,Y}\, \int^\infty_0 d t  e^{ -t^2\,Y}\frac{t^2}{t^2 + \bar{Q}^4_T}\,\,\nn\\
&& \Lb t = \sqrt{\omega_0 - \omega},  \bar{Q}_t \,=\,\sqrt{\alpha'_{eff}\pi \sqrt{D}}\,\, Q_T \Rb ~=\,\,
  \h \sqrt{D} \bar{Q}^2_T \,R^2_\pi\,\,e^{\omega_0\,\,Y}\,\,\Bigg\{ - e^{z^2} + \frac{1}{\sqrt{\pi}z} \,+\,e^{z^2} erf(z)\Bigg\}
   \eea

   where $z = \bar{Q}^2_T \,\sqrt{Y}$ and $erf(z)$ is the error function (see function $\Phi(z)$ in  Ref.\cite{RY}, formula  {\bf 8.25}).
   
   Since
\beq \label{QT2}
   \Big\{ \dots\Big\}\,\,\longrightarrow\,\,\left\{ \begin{array}{l l} z \,\ll\,1,\,\, & \,\,\,\frac{\pi}{z}\,\,-\,\,\pi \\
  &\\
z\,\gg\,1, &\,\,\,\frac{\sqrt{\pi}}{2\,\,z^3} \end{array}\right.
\eeq
                                    
   one can see that at small values of $Q_T$       $\mbox{Im}\,G$ behaves with the slope that has been calculated, and at large $Q_T$ in falls down as $1/\Lb Q^4_T\,\,Y^{3/2}\Rb$.                          
      \subsection{Dependence on the sizes of interacting dipoles}                            
      We have discussed the Green's function for the interaction of two dipoles with the same     large sizes $R_\pi$. Coming back to \eq{SCA} one can see that we need actually the Green's function for different dipole sizes. Let us consider first  $\mbox{Im}\, G_{\pom}\Lb Y, t = - Q^2_T; r,  R_\pi\Rb$. The typical processes which can be described by such Green's function include  inclusive deep inelastic scattering as well as  diffraction dissociation by the virtual photon.  We need to replace one (say, the upper in \fig{piloop}) BFKL Pomeron Green's function  by $G_\pom\Lb \omega, Q_T; r, R_\pi\Rb $ which is given by \eq{GBFKL}: viz.
          \beq \label{GDIS1}
 G_\pom\Lb \omega, Q_T; r,  R_\pi\Rb\,\,=\,\ \,\frac{\pi}{\kappa_0}\Bigg\{ \Lb \frac{R_\pi}{r}\Rb^{\kappa_0}\,\,+\,\, \Lb \frac{r}{R_\pi}\Rb^{\kappa_0}\,\,-\,\,2\,\Lb Q^2_T\,r\,R_\pi\Rb^{\kappa_0}\Bigg\}
  \eeq 
   As a result \eq{SOLY} should be replaced by the following equation
   \beq \label{GDIS2}  
   \mbox{Im}\, G_{\pom}\Lb Y, Q_T; r, R_\pi\Rb \,\,= \,\,r\, R_\pi\, \sqrt{D}\,  \int_C\!\frac{d\, \delta\omega}{2 \pi i}\,e^{\omega_0\,Y} \, e^{ \delta\omega\,Y} \frac{\Big\{ \Lb \frac{R_\pi}{r}\Rb^{\sqrt{\delta\omega/D}}\,\,+\,\, \Lb \frac{r}{R_\pi}\Rb^{\sqrt{\delta\omega/D}}\,\,-\,\,2\,\Lb Q^2_T\,r\,R_\pi\Rb^{\sqrt{\delta\omega/D}}\Big\}}{\sqrt{\delta\omega}\,\,+\,\,\alpha'_{eff}\pi \sqrt{D}\,Q^2_T}
     \eeq      
Closing contour $C$ in the integral over $\delta \omega$  as it is shown in\fig{cont}  and using the same notations as in \eq{QT1} we obtain
\bea \label{GDIS3}
 &&   \mbox{Im}\, G_{\pom}\Lb Y, Q_T; r, R_\pi\Rb\,\, = \,\,2\,r\, R_\pi\, \sqrt{D}\,e^{\omega_0 \,Y}\, \int^\infty_0 d t  \,e^{ -t^2\,Y}\\
    &&\times\,\,\Bigg\{ \frac{t^2}{t^2 + \bar{Q}^4_T} \Lb\cos\Lb \frac{t}{\sqrt{D}}\,\ln \Lb \frac{R_\pi}{r}\Rb \Rb\,\,-\,\,\cos\Lb  \frac{t}{\sqrt{D}}\,\ln\Lb Q^2_T \,r\,R_\pi   \Rb\Rb\Rb  \,\,-\,\,  \frac{t \bar{Q}^2_T}{t^2 + \bar{Q}^4_T} \,\sin\Lb\frac{t}{\sqrt{D}}\,\ln\Lb Q^2_T \,r\,R_\pi   \Rb\Rb \Bigg\}\nn
    \eea     
\subsubsection{\bf$\mathbf{Q_T}$ = 0}      

At $Q_T=0$ the integral over $t$ in \eq{GDIS3} can be taken and it leads to
\beq \label{GDIS4}
  \mbox{Im}\, G_{\pom}\Lb Y, Q_T; r, R_\pi\Rb\,\, = \,\,2\,r\, R_\pi\, \sqrt{\pi\,D}\,\exp\Lb\omega_0 \,Y\,\,-\,\,\frac{\ln^2\Lb R_\pi/r\Rb}{4\,D\,Y}\Rb\,\,\frac{\Lb 1 \,+\,erf\Lb i \frac{\ln\Lb R_\pi/r\Rb}{ 2 \sqrt{D\,Y}}\Rb\Rb}{2 \sqrt{Y}}
  \eeq
 One can see that for small $ \frac{\ln\Lb R_\pi/r\Rb}{ 2 \sqrt{D\,Y}} \,\ll\,1$  $   erf\Lb i \frac{\ln\Lb R_\pi/r\Rb}{ 2 \sqrt{D\,Y}}\Rb $ tends to zero and the integral in \eq{GDIS3} can be calculaterd using the steepest decent method with the saddle point in $t_{SP} = i \frac{\ln\Lb R_\pi/r\Rb}{ 2 \sqrt{D\,Y}}$.
 \subsubsection{\bf$\mathbf{ Q^2_T\,r\,R_\pi}\,\,\ll\,\,1$}
  In this kinematic region we can neglect the last term in \eq{GDIS1}. Taking the integral using the steepest decent method we obtain
  \bea \label{GDIS5}
    \mbox{Im}\, G_{\pom}\Lb Y, Q_T; r, R_\pi\Rb\,\, &=& \,\,2\,r\, R_\pi\, \sqrt{\pi\,D}\,\exp\Lb\omega_0 \,Y\,\,-\,\,\frac{\ln^2\Lb R_\pi/r\Rb}{4\,D\,Y}\Rb\,\,\frac{ t^2_{SP}}{t^2_{SP}\,\,+\,\,\bar{Q}_T^4}\,\,\nn\\
    &=&\,\,   \,\,2\,r\, R_\pi\, \sqrt{\pi\,D}\,\exp\Lb\omega_0 \,Y\,\,-\,\,\frac{\ln^2\Lb R_\pi/r\Rb}{4\,D\,Y}\Rb\,\,\frac{1}{1 - \bar{Q}^4_T \,4\,D\,Y/ \ln^2\Lb R_\pi/r\Rb}
    \eea
    We can trust \eq{GDIS5} for $Q^4_T\,\,\leq\,\,  \ln^2\Lb R_\pi/r\Rb/\Lb 4\,D\,Y  \Rb$. This inequality specifies the kinematic region in which the pion loops  gives an essential contribution.
     \subsubsection{\bf$\mathbf{ Q^2_T\,r\,R_\pi}\,\,\geq\,\,1$}    
     In this kinematic region we need to use all three terms of \eq{GDIS1}. Using the steepest decent method we obtain
     \bea \label{GDIS6}
     \mbox{Im}\, G_{\pom}\Lb Y, Q_T; r, R_\pi\Rb\,\, &=&       \,\,2\,r\, R_\pi\, \sqrt{\pi\,D}\,\exp\Lb\omega_0 \,Y\,\,-\,\,\frac{\ln^2\Lb R_\pi/r\Rb}{4\,D\,Y}\Rb\,\,\frac{1}{1 - \bar{Q}^4_T \,4\,D\,Y/ \ln^2\Lb R_\pi/r\Rb}\nn\\
     &+&   \,\,2\,r\, R_\pi\, \sqrt{\pi\,D}\,\exp\Lb\omega_0 \,Y\,\,-\,\,\frac{\ln^2\Lb Q^2_T\,r\,R_\pi\Rb}{4\,D\,Y}\Rb\,\,\frac{\bar{Q}^2_T \sqrt{\frac{\ln\Lb Q^2_T\,r\,R_\pi\Rb}{4\,D\,Y}}}{1 - \bar{Q}^4_T \,4\,D\,Y/ \ln^2\Lb Q^2_T\,r\,R_\pi \Rb}
     \eea
               \subsubsection{Impact parameter dependence}
               
               Using \eq{GDIS2} we can find the Green's function in $b$-representation using the general form of the BFKL Pomeron given by \eq{GDIS1}.  It takes the form
               \bea \label{GDIS7}
                  \mbox{Im}\, G_{\pom}\Lb Y, b; r, R_\pi\Rb \,\,=\,\,         
   \frac{2 \pi R^2_\pi}{\alpha'_{eff}}\, \int_C\!\!\! \delta\omega \,e^{\omega_0\,Y} \, e^{ \delta\omega\,Y}\!\!\!\!\!\!
  && \Lb \left\{ \Lb \frac{R_\pi}{r}\Rb^{\sqrt{\delta\omega/D}}\,\,+\,\, \Lb \frac{r}{R_\pi}\Rb^{\sqrt{\delta\omega/D}}\right\} \,K_0\Lb \sqrt{\frac{\sqrt{\delta \omega}}{\alpha'_{eff} \,D\,\pi}}\,\, b\Rb\right.\nn\\
  &&\,\,\left.\,~~~~-\,2\,\,\Lb   \frac{r\,R_\pi\,\sqrt{\delta \omega}}{\alpha'_{eff} \,D\,\pi}\Rb^{\sqrt{\delta\omega/D}}\, K_{\sqrt{\delta\omega/D}}\Lb \sqrt{\frac{\sqrt{\delta \omega}}{\alpha'_{eff} \,D\,\pi}}\,\, b\Rb\Rb
    \eea
   We can single out several kinematic regions where \eq{GDIS7} takes different form:
   \begin{enumerate}
   \item\quad $b$  is small and $R_\pi \,\gg\,1$. The main contribution stems from the saddle point $\sqrt{\delta \omega_{SP}}\,=\,- \ln\Lb R_\pi/r\Rb/\Lb 2 \sqrt{D} Y\Rb$ . If $r\,R_\pi \sqrt{\delta\omega_{SP} }\,\ll\,1$ we can neglect the second term in \eq{GDIS7} and the amplitude  will be equal to \eq{GDIS4} which is multiplied by $K_0\Lb \sqrt{\frac{\sqrt{\delta \omega_{SP}}}{\alpha'_{eff} \,D\,\pi}}\,\, b\Rb  $. The second term in \eq{GDIS7} can be neglected.
   \item  \quad $b$  is small and $R_\pi \,\gg\,1$, but $r\,R_\pi \sqrt{\delta\omega_{SP}} \,\gg\,1$.
   The second term should be taken into account with the saddle point in $\sqrt{\delta \omega_{SP}}\,=\,- \ln\Lb R_\pi\,r\Rb/\Lb 2 \sqrt{D} Y\Rb$.
      \item  \quad $b$  is large. The both terms can be calculated with the saddle point from \eq{LB2} leading to the same behaviour at large $b$ which is given by the exponent in \eq{LB3}. The estimates for how large have to be $b$, come from the condition $ \delta \omega_{SP}$ of \eq{LB2} is much larger than $ \, \ln\Lb R_\pi/r\Rb/\Lb 2 \sqrt{D} Y\Rb$.       
      \end{enumerate}      
      We hope that we have demonstrated how the pion loops affect the scattering amplitude,  leaving   more elaborate calculations to  the reader.
      
      \subsection{$\mathbf{b}$ dependence for general case of two dipoles with different sizes}
 In this section we consider the large $b$ behaviour of the Green's function for the general case of two dipoles with sizes $r_1$ and  $r_2$. For such scattering the first diagram in \fig{piloop}  looks differently than all others and  the sum of all diagrams can be written as         
   \bea \label{GC1}
 && G\Lb \omega, Q_T; r_1, r_2\Rb\,\,=\\
 &&G_\pom\Lb \omega, Q_T; r_1, r_2\Rb\,\,\,+\,\,\,  G_\pom\Lb \omega, Q_T; r_1, R_\pi \Rb   \Sigma\Lb \omega\Rb     \mathbf{G}\Lb \omega, Q_T; R_\pi, R_\pi\Rb  G^{-1}_\pom\Lb \omega, Q_T; R_\pi, R_\pi\Rb\,G_\pom\Lb \omega, Q_T; r_2, R_\pi \Rb   \nn         
  \eea         
 As it has been discussed for large impact parameter behaviour we can restrict ourselves by the region of very small $Q_T$ and neglect the last term in the BFKL Pomeron Green's function of \eq{GBFKL}.            
            
    Plugging in \eq{GC1} \eq{GBFKL} and     \eq{SOLMR} we can re-write \eq{GC1} in the form
    \bea \label{GC2}
  G\Lb \omega, Q_T; r_1, r_2\Rb\,\,&=&\,\,  - \frac{\pi \sqrt{D}}{\sqrt{\omega\,-\,\omega_0}}  \Bigg\{ \Lb \frac{R^2_\pi}{r_1\,r_2}\Rb^{\sqrt{(\omega - \omega_0)/D}}\,+\,    \Lb \frac{R^2_\pi}{r_1\,r_2}\Rb^{-\,\sqrt{(\omega - \omega_0)/D}}  \Bigg\}   \nn\\
  &+& \,\, \frac{\pi \sqrt{D}}{\sqrt{\omega\,-\,\omega_0}\,\,+\,\,\alpha'_{eff}\pi \sqrt{D}\,Q^2_T}
  \Bigg\{ \Lb \frac{r_1}{r_2}\Rb^{\sqrt{(\omega - \omega_0)/D}}\,+\,    \Lb \frac{r_1}{r_2}\Rb^{-\,\sqrt{(\omega - \omega_0)/D}}  \Bigg\}    \eea 
     
   The last term leads to the exponential fall down at large $b$ that has been discussed in section IV-B    (see \eq{SIG2} ).  The first term does not depend on $Q_T$ and gives $\delta \Lb b \Rb$  for $b$ dependence., and, therefore does not contribute  in  the scattering amplitude at large $b$.
     
 \section{Conclusions}
 In this paper we proposed a solution to the longstanding problem of the CGC/saturation approach: the power-like fall  of the scattering amplitudes at large $b$.  This decrease  leads to the violation of the Froissart theorem and  makes the approach theoretically inconsistent. We showed in the paper that sum of the pion loops  results in the exponential   fall of the scattering amplitude at large  impact parameters and in the restoration of the Froissart theorem. Hence we resolve the fundamental difficulty of the CGC/saturation approach.
 
 Our solution is based on the assumption which we have to make in the absence     of  consistent approach to non-perturbative QCD.  We assume that the BFKL Pomeron which stems from perturbative QCD calculations, can be used for the description of the long distance physics but the typical distances in this Pomeron is  smaller that $1/m_\pi$. Making this assumption more specific we  assume that the BFKL Pomeron, analytically continuated to the $t$-channel, has all singularities in the $t$-channel  at the values of  $t$ which are much larger than $4 m^2_\pi$. We see support of this assumption in the lattice calculations  which show that the Pomeron trajectory\cite{GLUETR} has the lightest glueball on it with $m^2_G = 5 \,GeV^2$.
 
Our summation procedure preserves the spectrum of the BFKL Pomeron which depends on the short distances as has been illustrated in the numerous model 
 approaches in which a new dimensional scale have been introduced(see Refs,\cite{FIIM,LERYB1,LERYB2,GBS1,GKLMN,HAMU,MUMU,BEST1,BEST2,KOLE,LETA,LLS}).
 
 We believe, that our solution  will encourage experts to look  for a new dimensional scale with more microscopic origin than the pion exchange.

  \section{Acknowledgements}
  
 We thank our    colleagues at Stony Brook university, Brookhaven lab., Tel Aviv university and UTFSM for encouraging discussions.   This research was supported by the BSF grant 2012124  and by the  Fondecyt (Chile) grant 1140842.

 \end{document}